\documentclass[12pt]{iopart}

\usepackage{iopams}

\usepackage{bm}
\usepackage{graphicx}

\newcommand{\vecr}{\mathbf{r}}
\newcommand{\vecR}{\mathbf{R}}

\begin{document}

\title[NP-hardness of cluster minimization]{NP-hardness of the cluster minimization problem revisited}

\author{Artur B. Adib}

\ead{artur@brown.edu}

\address{Physics Department, Brown University, Providence, RI 02912, USA}

\date{\today}

\begin{abstract}
The computational complexity of the ``cluster minimization problem'' is revisited [L. T. Wille and J. Vennik, J. Phys. A {\bf 18}, L419 (1985)]. It is argued that the original NP-hardness proof does not apply to pairwise potentials of physical interest, such as those that depend on the geometric distance between the particles. A geometric analog of the original problem is formulated, and a new proof for such potentials is provided by polynomial time transformation from the independent set problem for unit disk graphs. Limitations of this formulation are pointed out, and new subproblems that bear more direct consequences to the numerical study of clusters are suggested.
\end{abstract}



\maketitle

The present contribution addresses the inherent computational difficulty of finding the ground state configuration $\vecR = (\vecr_1,\ldots,\vecr_N)$ of a classical $N$-particle system with total potential energy
\begin{equation} \label{U}
  U(\vecR) = \sum_{i=1}^N \sum_{j>i}^N u(|\vecr_j - \vecr_i|),
\end{equation}
where each $\vecr_i \in \mathbb{R}^3$. Though fairly general in scope, this problem is of particular interest in the characterization of the energy landscapes of atomic clusters (see e.g. \cite{wales99,wales03}), and bears important consequences to the efficiency of Monte Carlo sampling methods for clusters at low temperatures.

When addressing such fundamental aspects of optimization problems, a primary goal is to establish whether the given problem is ``NP-hard,'' a complexity class also known as {\em intractable} since their solution in the worst-case scenario requires a number of computational steps that increases exponentially with the ``size'' of the problem (intractability hinges upon the validity of the P $\neq$ NP conjecture, which is nevertheless widely accepted, see \cite{garey-johnson79,sipser97} for details, or \cite{mertens02} for a gentle introduction aimed at the physics audience). Establishing the membership of a problem in the NP-hard class is, on the one hand, of fundamental importance to the field of computational complexity, for a given polynomial time solution to the problem would collapse the entire NP class into P, and on the other hand of practical interest to the problem itself, for it indicates that the chances of finding a general-purpose algorithm with polynomial efficiency across all instances of the problem are very small, thereby encouraging the development of specialized algorithms that can benefit from whatever physical insight one has into the specific instance of the problem at hand. Proofs of NP-hardness are especially welcome in light of recent efforts to characterize the boundaries between exponential and polynomial behavior in ``typical'' instances of NP-complete problems (see e.g. \cite{anderson99,monasson99,mezard02}), an endeavor that has the potential of providing valuable guidance for dealing with NP-hard optimization problems.

A study on the computational complexity of classical systems interacting via pairwise potentials has been reported in Ref.~\cite{wille85}, a celebrated work that is frequently cited in the literature as providing a proof that the ``cluster minimization problem'' is NP-hard (see e.g. \cite{stillinger90,yi91,maranas92,deaven95,wales96,doye98}). It is part of the goal of the present work to argue that this proof is too general, and that a different proof is necessary for physically relevant potentials such as those described by Eq.~(\ref{U}) (i.e. those that depend on the geometric distance between the particles), and hence that the NP-hardness of the ``physical'' cluster minimization problem is still an open issue. The remainder of this paper is thus divided in three parts: a critical analysis of Ref.~\cite{wille85}, a proof of NP-hardness valid for potentials of the form of Eq.~(\ref{U}), and a discussion on the limitations of this proof.

One of the simplest ways of proving that a given problem $\pi$ is NP-hard is by ``restriction,'' i.e. by showing that a specific instance of the problem $\pi$ coincides with a problem $\pi'$ already known to be NP-hard (much of the present discussion on computational complexity was adapted from \cite{garey-johnson79}). Often times, however, the connection between the problem of interest and a known intractable problem is not immediate; in this case, one can still establish the NP-hardness of $\pi$ by ``polynomial transformation'' from $\pi'$, i.e. by showing that every instance $I'$ of $\pi'$ can be transformed to an instance $I=f(I')$ of $\pi$ in polynomial time, and that a solution of $I$ also solves $I'$. The NP-hardness proof of Wille and Vennik was based on such a transformation \cite{wille85}: Formulating the cluster minimization problem in terms of ``graphs'' (see e.g. \cite{sipser97}), these authors considered what will be henceforth referred to as the {\tt WEIGHTED EDGE} problem, showing that the {\tt TRAVELING SALESPERSON} ({\tt TSP}) problem can be transformed to {\tt WEIGHTED EDGE}; since {\tt TSP} is known to be NP-hard, so must be {\tt WEIGHTED EDGE}. In the following, I will analyze the proof offered by these authors, arguing that it requires pair ``potentials'' that are far more general than those of physical interest, which leaves the possibility that the case of more restrictive and physically relevant potentials may be treated by a polynomial-time algorithm.

\begin{figure}
\begin{center}
\includegraphics[width=3.3in]{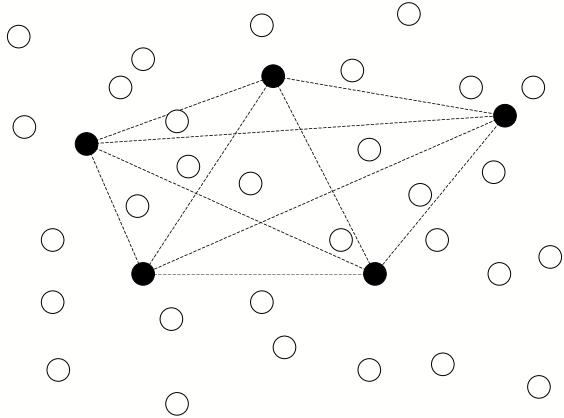} 
\caption{Illustration of the {\tt WEIGHTED EDGE} problem, with vertices represented by circles and edges by dotted lines. The problem is to choose $N$ out of all the vertices such that the sum of the corresponding edge weights $w(e)$ is minimal. Although the entire graph is complete, for clarity of illustration only edges connecting the vertices of a hypothetical solution with $N=5$ are shown.}
\label{fig:weighted}
\end{center}
\end{figure}

First, let us state Wille and Vennik's problem following the standard format of Garey and Johnson \cite{garey-johnson79}:

\begin{quote}
{\tt [WEIGHTED EDGE]} {\em Instance:} A complete graph $G=(V,E)$, a weight function $w(e)$ for each $e \in E$, and a number $N<|V|$. {\em Problem:} Find a subgraph $G'=(V',E')$ of size $|V'|=N$ such that
\begin{equation*}
  \sum_{e \in E'} w(e) \,\, \textrm{is minimal.}
\end{equation*}
\end{quote}
Here a complete graph $G=(V,E)$, where $V$ are the vertices and $E$ are the edges connecting the vertices, is one in which every pair of vertices is connected by an edge (i.e. a ``clique''), and the notation $|V|$ indicates the number of vertices. An illustration of this problem is provided in Fig.~\ref{fig:weighted}.

When formulating the cluster minimization problem as {\tt WEIGHTED EDGE}, the authors of Ref.~\cite{wille85} had in mind the following analogy: $V$ is the set of ``sites'' that the $N$ particles are allowed to occupy, and for every pair of sites with edge $e=(v_i,v_j)$, the function $w(e)=w(v_i,v_j)$ measures the strength of the interaction between two hypothetical particles placed in these sites. The set $V$ is of course discrete, reflecting the discretization of the problem that accompanies any computational method, but notice that no specific structure is given to either $V$ or $w(e)$ (this arbitrariness will be discussed later in the text).

As already mentioned, the proof that {\tt WEIGHTED EDGE} is NP-hard was provided in Ref.~\cite{wille85} by showing that it ``contains'' {\tt TSP}, more precisely by showing that there exists a weight function $w(e)$, a choice of $|V|$ as a function of $N$, and a labeling of the vertices in terms of pairs of ``cities,'' such that a solution of {\tt WEIGHTED EDGE} would yield a minimum {\tt TSP} tour.\footnote{A more direct proof that {\tt WEIGHTED EDGE} is NP-hard can be given. Indeed, given any graph $G^*=(V^*,E^*)$, construct a complete graph $G=(V^*,E)$, setting $w(v_i^*,v_j^*)=1$ if there is a corresponding $(v_i^*,v_j^*) \in E^*$, and $w(v_i^*,v_j^*)=2$ otherwise. The solution of {\tt WEIGHTED EDGE} for the graph thus constructed has total weight equal to $N(N-1)/2$ if and only if $G^*$ contains a clique of size $N$, and the foregoing transformation can be done in polynomial time. But determining if a graph contains a clique of a given size is known to be an NP-complete problem ({\tt CLIQUE}, see \cite{garey-johnson79}), and thus {\tt WEIGHTED EDGE} must be NP-hard, q.e.d.} Though not stated in \cite{wille85}, it is easy to check that this labeling and choice of $w(e)$ can be done in polynomial time given any instance of {\tt TSP}. One additional observation not explicitly mentioned by Wille and Vennik is that, since the NP-hardness of {\tt TSP} means that worst-case solutions to the problem are found in time exponential in the number of {\em cities}, their result implies only that worst-case solutions to {\tt WEIGHTED EDGE} are found in time exponential in the number of {\em sites}, not in the number of particles $N$ as one would like to show (recall that the general instance of {\tt WEIGHTED EDGE} does not tie $|V|$ to $N$). Nonetheless, it stands to reason that any algorithm relevant to the cluster minimization problem will adjust the size of $V$ so as to ensure comparable spatial resolutions for clusters of different sizes, and a reasonable choice would be a polynomial in $N$, i.e. $|V| \propto N^p$ for some integer $p > 0$. Thus, the NP-hardness of {\tt WEIGHTED EDGE} with this additional property says that the completion time of any algorithm that solves the problem will indeed scale exponentially with the number of particles $N$.

Finally, observe that the choice of $w(e)$ that allows the connection with {\tt TSP} (Eq.~(4) in Ref.~\cite{wille85}) is too general to reflect physical problems of interest; in particular, it requires the possibility that the ``potential'' between two sites depends on {\em each individual site} in question, and not just on their relative geometric distance. Although the authors state that their proof remains true for spherically symmetric potentials ``by comparing it to the general and the Euclidean traveling salesperson problem'' \cite{wille85}, this is far from clear: It is difficult to imagine a polynomial time transformation from a generic instance of the (Euclidean or not) {\tt TSP} problem to an instance of the {\em geometric }{\tt WEIGHTED EDGE} problem, where the vertices are points in the Euclidean space and the weights $w(e)$ are geometric distances (see {\tt CLUSTER MINIMIZATION} below). The proof of the Euclidean {\tt TSP} that the authors refer to is quite elaborate, and requires methods other than simple restriction \cite{garey76}, and it is also not clear how this proof generalizes to their problem. In the following, an explicit proof for potentials that depend on the geometric distance between the particles will be provided.

The problem of minimizing the energy of $N$ particles with spherically symmetric pairwise interaction can be stated as:
\begin{quote}
{\tt [CLUSTER MINIMIZATION]} {\em Instance:} A set of sites $S$ in $\mathbb{R}^3$, a number of particles $N < |S|$, and a potential function $u(r)$. {\em Problem:} Find a subset $S'=\{ \vecr_1, \ldots, \vecr_N \}$ of $S$ with size $|S'|=N$, such that
\begin{equation*}
  \sum_{i=1}^N \sum_{j>i}^N u(|\vecr_j - \vecr_i|) \,\,\, \textrm{is minimal.}
\end{equation*}
\end{quote}
Before we proceed, note the crucial difference with respect to {\tt WEIGHTED EDGE}: In {\tt CLUSTER MINIMIZATION}, it is the {\em relative geometric position} of the sites that determines the strength of the interaction potential. The proof of Ref.~\cite{wille85} for the {\tt WEIGHTED EDGE} problem does not apply here, and likewise there are several subproblems of {\tt WEIGHTED EDGE} not amenable to that proof which are nevertheless NP-hard by restriction to {\tt CLUSTER MINIMIZATION} (e.g. the case where the interaction depends on the vectorial distance, $u=u(\vecr_j-\vecr_i)$, can clearly be restricted to spherically symmetric potentials).

\begin{figure}
\begin{center}
\includegraphics[width=3.3in]{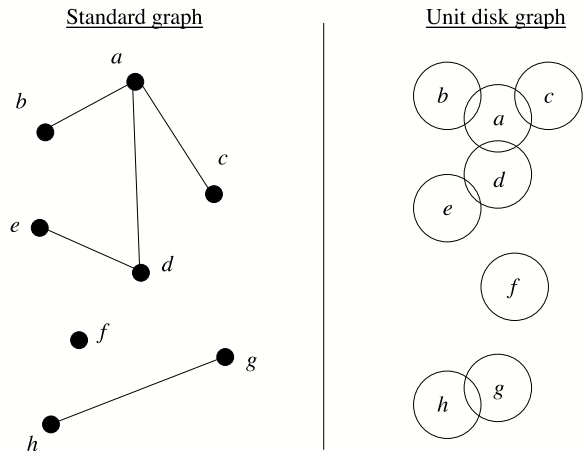} 
\caption{Illustration of a unit disk graph. Left: a graph drawn in the ``standard'' format; Right: the corresponding UDG. Note that the problem of determining whether an arbitrary graph can be represented as a UDG is itself NP-hard \cite{breu98}.}
\label{fig:disk}
\end{center}
\end{figure}

It will now be shown that the unit disk graph (UDG) problem {\tt UDG INDEPENDENT SET}, a decision problem known to be NP-complete \cite{clark90}, can be transformed to {\tt CLUSTER MINIMIZATION} in polynomial time. The problem can be stated as:
\begin{quote}
{\tt [UDG INDEPENDENT SET]} {\em Instance:} A set $D$ of disks of unit radius in $\mathbb{R}^2$, and an integer $K < |D|$. {\em Problem:} Does $D$ contain an independent set of size $K$, i.e. a subset $D'$ of $D$ with $K$ disks such that the disks never overlap with each other?
\end{quote}
The problem {\tt UDG INDEPENDENT SET} is one among various graph-theoretic problems that remain intractable when the graphs are restricted to be of the unit disk type, i.e. when vertices can be represented by disk centers, and edges by causing the corresponding disks to overlap in space \cite{clark90} (see illustration in Fig.~\ref{fig:disk}). The problem of determining whether an arbitrary graph can be represented by a unit disk graph is also NP-complete \cite{breu98}, and so far the only interesting problem in this context that allows a polynomial time solution seems to be {\tt UDG CLIQUE} \cite{clark90}.

The proof proceeds by choosing the sites $S$ of {\tt CLUSTER MINIMIZATION} to coincide with the center of the disks $D$ of a given instance of {\tt UDG INDEPENDENT SET} (this can clearly be done in polynomial time, and effectively places the sites in a plane), and setting
\begin{equation*}
  u(r) = \cases{
			2, & for $(r\leq 1)$ \\
		    1. & for $(r>1)$
		 }
\end{equation*}
Therefore, this instance of {\tt CLUSTER MINIMIZATION} has a solution with total energy equal to $N(N-1)/2$ if and only if $D$ has an independent set with $N$ disks. In other words, an answer to {\tt UDG INDEPENDENT SET} is immediately available from a solution of the corresponding {\tt CLUSTER MINIMIZATION} problem (recall that the evaluation of the energy of an $N$-particle system interacting with pairwise potentials can be done in polynomial time, more explicitly in $\mathcal{O}(N^2)$ steps). Thus, since {\tt UDG INDEPENDENT SET} is NP-complete, it must be that {\tt CLUSTER MINIMIZATION} is an NP-hard problem, q.e.d. (Note the similarity of this argument to that used to show that {\tt HAMILTONIAN CIRCUIT} can be transformed to {\tt TSP} in \cite{garey-johnson79}, pp. 35-36).

The above formulation of the cluster minimization problem is subject to some caveats. As already discussed in the case of {\tt WEIGHTED EDGE}, the assumption that the total number of sites $|S|$ grows with the number of particles $N$ is required in order to have an intractability result in terms of $N$. A more subtle limitation of the present formulation can be appreciated by stating the foregoing proof in words: {\em Given an arbitrary set of sites in $\mathbb{R}^3$ and a spherically symmetric but otherwise arbitrary pair-potential energy function $u(r)$, no algorithm can find the minimum energy configuration of the system in time bounded by a polynomial in the number of particles $N$}. (As usual, this result hinges upon the P $\neq$ NP conjecture, see the introductory paragraphs above). This is perhaps not the most relevant result for the cluster minimization problem, since in practice one seldom uses a discretization different from the simple cubic lattice (this happens e.g. when the components of $\vecr$ are represented by finite-precision, binary numbers). Ideally, one would like to show that for any ``reasonable'' discrete representation of the continuum (see e.g. \cite{keil92} for a formalization and examples of this concept), the arbitrariness in the choice of $u(r)$ by itself is sufficient to make the problem NP-hard, i.e. that for any such lattice, there are choices of $u(r)$ leading to intractable solutions. The foregoing proof of course fails when the sites are arranged in a simple cubic structure, for the existence of an independent set of a given size in this case is straightforwardly determined by the radius of the disks and the lattice spacing. In fact, it is known that a polynomial time algorithm exists for finding independent sets in unit disk graphs restricted to ``grid'' structures \cite{clark90}. The NP-hardness of this subproblem seems considerably more difficult to prove than the more general {\tt CLUSTER MINIMIZATION} problem introduced in this study, and would certainly be a welcome result in the literature.

In summary, a critique of the celebrated NP-hardness proof of the cluster minimization problem \cite{wille85} was offered, showing that a separate proof is necessary when the pairwise potential is a function of the geometric distance between the particles only. Using a geometric generalization of the formulation adopted in Ref.~\cite{wille85}, an independent proof for such cases was presented by transformation from the independent set problem for unit disk graphs \cite{clark90}. Limitations of this new proof were discussed in the last paragraph, and a subproblem that fixes the lattice structure was suggested as being more relevant to traditional numerical implementations of the cluster minimization problem. To this author's knowledge, however, the intractability of this subproblem remains unknown.

\ack

I am indebted to Heather Partner for introducing me to the subject of computational complexity and reading the original manuscript. Discussions with Jimmie Doll and Tongsik Lee are also greatly appreciated. This research was supported in part by the U.S. Department of Energy under grant DE-FG02-03ER46074.

\end{document}